# COVID_MTNet: COVID-19 Detection with Multi-Task Deep Learning Approaches

Md Zahangir Alom, M M Shaifur Rahman, Mst Shamima Nasrin, Tarek M. Taha, and Vijayan K. Asari
Department of Electrical and Computer Engineering, University of Dayton, Dayton, OH, USA
Emails: {alomm1, rahmanm24, nasrinm1, ttaha1, vasari1}@udayton.edu

## Abstract

*COVID-19 is currently one the most life-threatening problems around the world. The fast and accurate detection of the COVID-19 infection is essential to identify, take better decisions and ensure treatment for the patients which will help save their lives. In this paper, we propose a fast and efficient way to identify COVID-19 patients with multi-task deep learning (DL) methods. Both X-ray and CT scan images are considered to evaluate the proposed technique. We employ our Inception Residual Recurrent Convolutional Neural Network with Transfer Learning (TL) approach for COVID-19 detection and our NABLA-N network model for segmenting the regions infected by COVID-19. The detection model shows around 84.67% testing accuracy from X-ray images and 98.78% accuracy in CT-images. A novel quantitative analysis strategy is also proposed in this paper to determine the percentage of infected regions in X-ray and CT images. The qualitative and quantitative results demonstrate promising results for COVID-19 detection and infected region localization.*

## 1. Introduction

Coronavirus Disease 2019 (COVID-19) is an infectious disease caused by a new virus that has not been previously identified. The disease causes respiratory illness with symptoms such as cough, fever, and in more severe cases, difficulty in breathing. . The COVID-19 family is causing thousands of deaths per day around the world. Even though this virus first started in Wuhan, China in December 2019, this dangerous virus is spreading around the world so quickly and more than 1.6 million people around 188 countries have been infected as of today when we write this article. Due to coronavirus, more than 100 thousand people have already died worldwide, and the number of new deaths has been increasing rapidly day by day. There is no treatment specifically approved for this virus so far. Thus, the World Health Organization (WHO) has declared COVID-19 as a pandemic disease. The mutation of this virus moving faster than anything and more deadly too. This is one of the fastest progressive disease ever seen before [2]. Since it is a new type of virus and it changes formation quickly, there is no specific guideline for the assessment or examination process and proper treatment. According to findings from Wuhan, Hubei province in China, isolation of infected patients through home quarantining is the best way to protect people not already infected with this deadly virus [2]. Thus, patients with COVID-19 symptoms must remain isolated and wear masks when near others. For primary examination, a sample of nasopharyngeal exudate is taken to test reverse transcription polymerase chain reaction (RT-PCR) followed by a chest X-ray. If the chest X-ray film is normal, then the patient can go home and take rest. If COVID-19 is confirmed, the patient must be admitted to a hospital. At the initial stages, COVID-19 affects both lungs, particularly the lower lobes, and especially the posterior segments, with a fundamentally peripheral and subpleural distribution. The chest CT is used for detecting COVID-19 in practically 50% of patients in the first two days. A researcher from Tongji Hospital in Wuhan, China, has concluded that the CT should be used as the primary screening or diagnostic method to identify COVID-19.

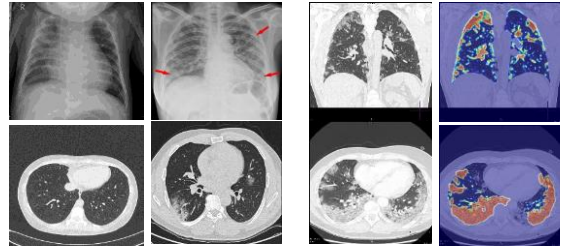

Figure 1: Example X-ray and CT images of normal and COVID-19 cases, and outputs of detection and infected region localization with NABLA-3 network.

Even though the RT-PCR has been considered as the gold standard for SARS-CoV-2 diagnosis, due to the limited supply and strict requirements for the laboratory environment, it delays the accurate diagnosis of patients. Hence, it is very difficult to pause the unexpected spreading of infection of COVID-19 diseases. By combining the clinical symptoms and signs, the chest computed tomography (CT) is a faster and easier approach to identify COVID-19 in clinical practices. According to doctors and researchers in China, CT image analysis is the key method to examine suspected patients for COVID-19 confirmation.

Recent reports also support this phenomenon [2,3]. It is crucial to detect COVID-19 infections at an early stage for suspected patients. It is also important for patient prognosis, control of this epidemic, and public health security. As the rate of newly affected global regions is increasing every day, and the number of new patients is increasing geometrically in those areas, it has become a difficult or sometimes impossible task to provide necessary treatment for them [3,4,5]. Hence, the availability of an Artificial Intelligence (AI) system can be helpful to identify the COVID-19 patients quickly and accurately using X-ray and CT images as shown in Figure 1.

The main motivation of this research is accurate and faster detection of COVID-19 patients as it is very important for the prevention and control of this pandemic. Most importantly, it can protect human lives around the world. We also present deep learning-based radiology image analysis methods that could provide state-of-the-art testing accuracy by far compared to existing methods [6]. Recently, several Deep Learning (DL) based methods have been presented and shown to be highly accurate for COVID-19 detection from CT images [7,12,17].

In this study, we apply advanced deep learning methods to investigate both X-ray and CT manifestations of coronavirus (COVID-19) pneumonia. These include classification and segmentation models for analysis of samples from two different modalities for detection and localization. For chest X-ray image analysis, due to the scarcity of publicly available COVID-19 X-ray samples, we have trained our model with a pneumonia dataset, and then utilized a Transfer Learning (TL) method for retraining with samples of COVID-19. The contribution of this paper can be summarized as follows:

- It presents an end-to-end COVID-19 detection and infected region localization method using deep learning approaches.
- The proposed methods are evaluated for both X-ray and CT images and achieved promising results for COVID-19 detection and infected region localization tasks.
- Training and validation are performed on publicly available samples which are collected from different sources around the world.

The contents of this paper are organized as the following. Section 2 discusses related works. Section 3 presents the proposed method and the models used in the implementation. The experimental setup is discussed in Section 5. Sections 6, 7, and 8 present detailed discussion on results, limitations of the proposed method, and conclusions respectively.

## 2. Related works

There are several AI systems that have been proposed for COVID-19 detection from X-ray and CT images. Some studies demonstrate the efficiency of X-ray and CT scan image analysis for detecting the COVID-19. Several researchers presented artificial Intelligence (AI) and image analysis-based methods for COVID-19 detection. In most of the cases, the Deep Learning based approaches have been applied and achieved very promising detection accuracy for COVID-19.

A detailed study has been conducted in [8] to illustrate the importance of early detection and management of COVID-19 patients. A literature review was published recently in [9] and claimed that the ground-glass and consolidative opacities on CT are sometimes undetectable on chest radiography. The study has also suggested that CT is a more sensitive modality of medical imaging. The COVID-19 infection pulmonary manifestation is predominantly characterized by ground-glass opacification with occasional consolidation on CT. The pros and cons of using X-ray and CT image analysis and its effectiveness for the screening of COVID-19 patients were explained in recent studies. In most of the cases, chest X-ray is considered as the primary screening method. In cases of patients without predominant disease, COVID-19 can be defined by chest X-ray. Otherwise the CT was recommended for determining COVID-19 infection. Another study demonstrated that chest CT has a pivotal role for the diagnosis and assessment of lung involvement in COVID-19 pneumonia [10]. CT plays a central role in the diagnosis of COVID-19 pneumonia and the decision has been taken by conducting an evaluation with 366 CT scans which were reviewed by two groups of radiologists. The objective of this study was to define disease progression and recovery of the illness and found that the peak period during illness were days 6-11 [11]. The performance of the radiologists has been analyzed and observed that the radiologist has high specificity but moderate sensitivity in differentiating COVID-19 from viral pneumonia on CT scan [12]. For COVID-19 detection, a CT scan image analysis method found that in 20 (56%) patients out of 36 patients about 2 days after symptom onset had normal [13]. Furthermore, for critical cases, images from both modalities are recommended for decision making.

Meanwhile, there are several DL based systems that have been proposed for X-ray and CT images in the last few months for COVID-19 detection. A DL based method has been proposed for COVID-19 detection from chest X-ray which was named COVID-19 Detection Neural Network (COVNet). The ReseNet50 was used as the backbone for COVNet. This model was tested on 4356 chest CTs from 3322 patients and it showed 0.96 and 0.95 scores for Area Under Curve and community acquired pneumonia (CAP) score respectively [14]. The study also demonstrated CT findings and compared for asymptomatic and symptomatic patients for COVID-19 detection and the results showed that there were no significant differences in age, sex distribution, or comorbidities for symptomatic and asymptomatic cases [14]. One of the most relevant and recent models, COVID-Net [15]], used a deep neural network with high architectural diversity and selective long-range connectivity. This model experimented on two open access data repositories, one for COVID-19 detection dataset and another for pneumonia detection dataset. The experimental results showed around 83.5% testing accuracy. In this case, both databases were merged for performing training and testing of the proposed method [15]. However, in our proposed approach, we have trained a model with a pneumonia dataset and used Transfer Learning (TL) for training the same model with a COVID-19 dataset.

A deep learning-based detection for COVID-19 from chest 3D CT volumes with weak labeled samples has been proposed recently. For each patient, a pretrained UNet model was applied for 3D lung region segmentation and then a 3D-CNN model was applied for predicting the probability of COVID-19 infections. The model was trained on 499 CT volumes and tested on 131 CT volumes and obtained 0.959 ROC AUC with 0.907 and 0.911 sensitivity and specificity respectively. Overall, this model showed around 90% accuracy [16]. In addition, another recent study showed very promising accuracy for pneumonia detection tasks where the model grouped three different categories: COVID-19, Influenza-A viral pneumonia, and healthy cases from CT images. The VNet based VNET-IR-RPN is used for region of interest segmentation for pulmonary tuberculosis. This deep learning-based method showed around 86.7% testing accuracy from CT images [17].

According to various studies presented in the literature, the X-ray and CT scans have been used quite often to recognize or identify the patients with COVID-19. Hence, we present here an end-to-end deep learning-based system for COVID-19 detection and infected region localization from both X-ray and CT images.

## 3. Methodology

In this implementation, we have used multiple models for different tasks where the classification model is employed for COVID-19 detection and the segmentation model is used for Region of Interest (ROI) detection for COVID-19. Our Inception Recurrent Residual Neural Network (IRRCNN) model is used for the COVID-19 detection task [18]. Our NABLA-N network is applied for infected region segmentation from X-ray and CT images [19].

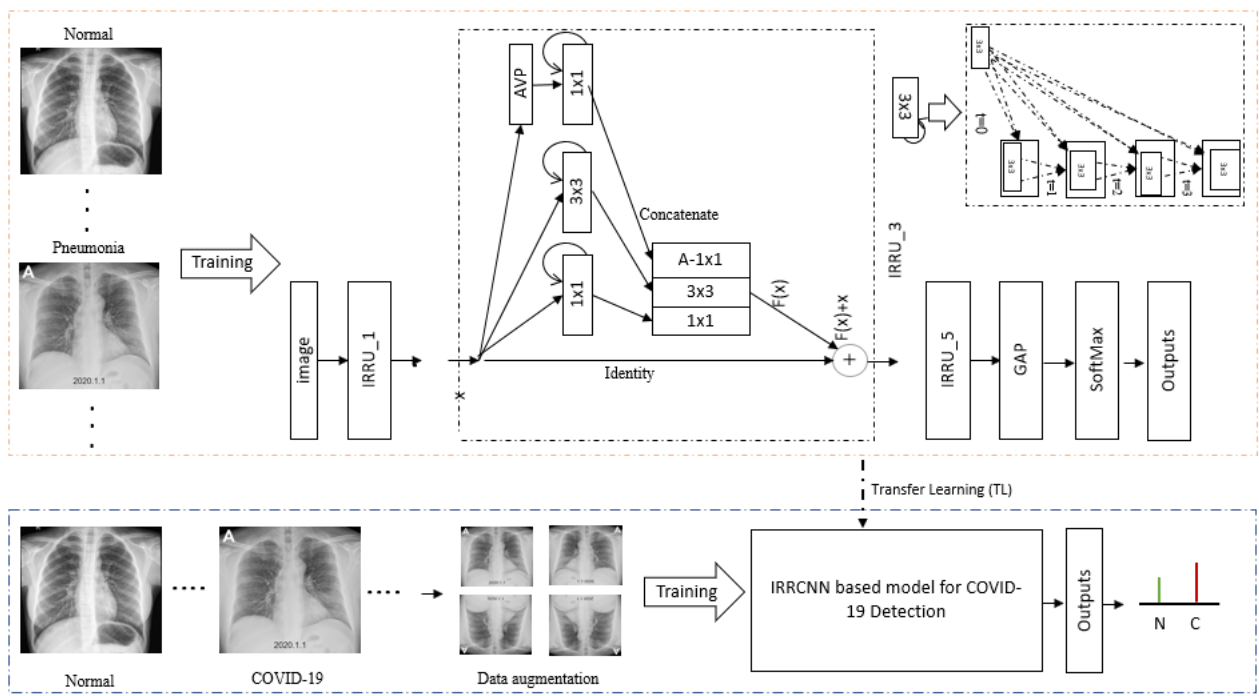

Figure 2: COVID-Det_X-ray: End-to-end system for pneumonia detection in the top row and COVID-19 detection with TL learning from pneumonia detection at the bottom. Bottom row shows the training phase for COVID-19 which include inputs, data augmentation, training model with TL and outputs (N: normal and C: COVID-19).

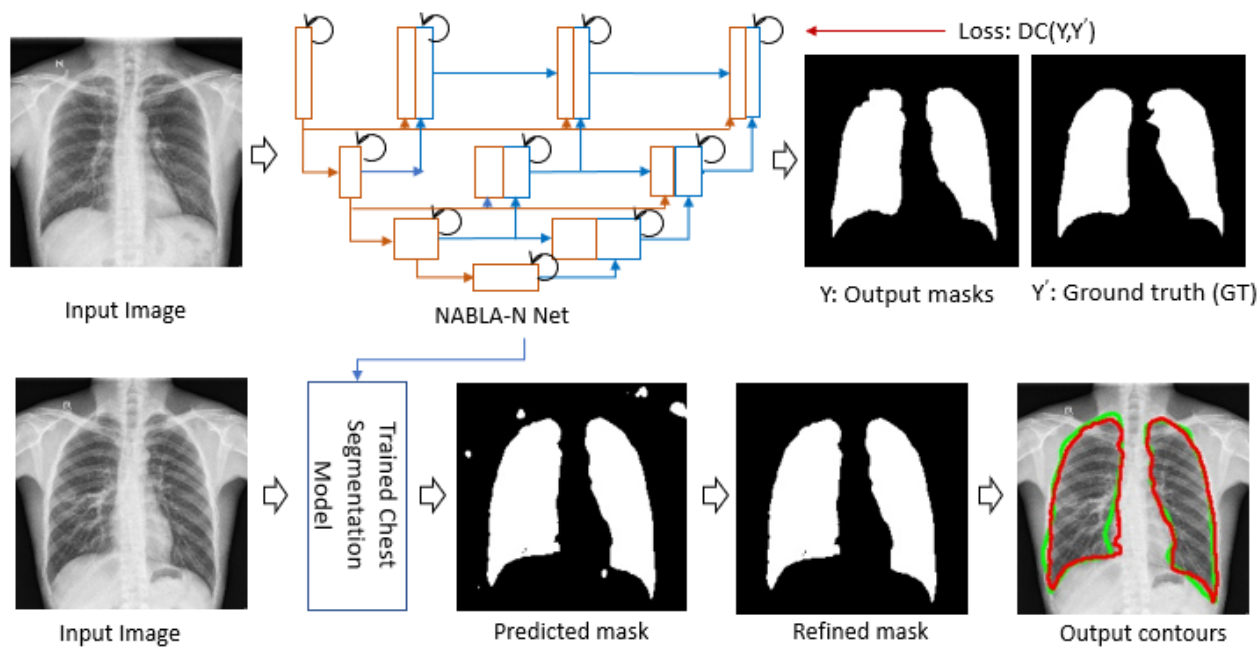

Figure 3: Chest segmentation model for X-ray (Chest-Seg_X-ray): the top row of the figure shows the training method and bottom row demonstrates the different steps for testing phase.

**COVID-19 Detection from X-ray-Image (COVID-Det_X-ray):** Firstly, we have used the IRRCNN model to classify the normal and pneumonia detection from chest X-ray images. Then, Transfer Learning (TL) method is used to retrain the model with samples to distinguish COVID-19 versus normal images in Chest X-ray images. The end-to-end training pipeline is shown in Figure 2. The IRRCNN model is used for this implementation where five Inception Recurrent Residual Units (IRRUs) are used [18]. The IRRUs are shown for unit 3 in the top of Figure 2. After successfully training the model for COVID-19 detection, the system has been tested with completely new samples collected from new patients.

**COVID-19 segmentation from X-ray-Images (COVID-Seg_X-ray):** Another system is developed for detecting infected regions due to COVID-19 in X-ray images. In this case, the NABLA-N model is applied for only chest region segmentation [19]. The overall framework of the end to end training and testing phases for chest segmentation method (Chest-Seg_X-ray) is shown in Figure 3. Mathematical morphological approaches are applied for performing the refinement and selecting appropriate contours for chest region extraction, and is shown in the last column of the second row in Figure 3. In the testing phase, precise chest regions are extracted. After generating segmentation masks for the chest regions, the mask is used to extract only the chest regions from the input images as shown in Figure 4(a). A classical image processing method and an adaptive thresholding approach are applied for extracting the features to identify the infected regions of COVID-19 from the segmented chest region in Figure 4(b). Finally, the output heatmap image with COVID-19 infected regions are shown in Figure 4(c).

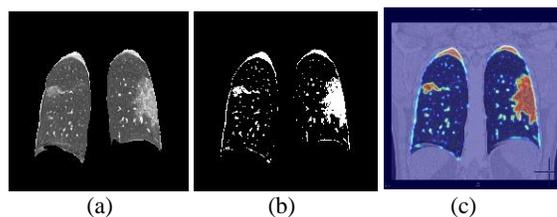

(a)          (b)          (c)

Figure 4: The pipeline processing for COVID-19 infected region detection: (a) chest regions, (d) mask for COVID-19 infected regions and (c) final heatmap image.

**COVID-19 detection from CT scans (COVID-Det_CT):** CT scans are found to be more effective to detect the infection of COVID-19 within a couple of days after infection. This is stated based on the assessment system followed by the experts in Wuhan, China. For COVID-Det_CT, we use our IRRCNN classification model, which is trained by datasets collected by us. The database samples are collected from normal CT and CT scans with COVID-19. From this implementation, the same strategy is employed as described for COVID-Det_X-ray.

**COVID-19 segmentation from CT scans (COVID-Seg_CT):** For mining the specific infected region affected by COVID-19 virus in lungs, we have developed an end to end system for segmentation of lung region from 2D images where the same NABLA-N model is used as mentioned in COVID-Seg_X-ray. First, we have trained the model on publicly available 2D lung images, Lung_Seg_CT. Then the trained Lung_Seg_CT model is directly applied to segment the lung region from the entire set of CT images. Classical image processing and morphological analysis are then applied to extract infected regions due to COVID-19. The entire system is named "COVID-Seg_CT". For the testing phase, the same processing pipeline as demonstrated for X-ray and shown in Figures 3 and 4 is used.

**Architecture details:** The IRRCNN model consists of an input layer, five IRRUs, a Global Average Pooling (GAP) layer, and a Softmax output layer. For this model, we have applied 1×1, and 3×3 kernels for IRRU. The entire model utilizes around 34M network parameters. On the other hand, the NABLA-N network consists of an encoding and three decoding units. The model architecture is as follows: 3→16×(3×3) →32×(3×3) →64×(3×3) →128×(3×3) →256×(3×3) →512×(3×3) →256×(3×3) →128×(3×3) →64×(3×3) →32×(3×3) →16×(3×3) →1. The general notation of $F_N \times (M \times N)$ where $F_N$ represents the number of feature maps and (M×N) represents the kernel size which is (3×3) kernels used in each layer except the last layer. The Rectified Linear Unit (ReLU) activation function is used all through the network in this implementation. At the end, a 1×1 convolutional layer is used to reduce the dimension of the feature maps to the single channel outputs with a sigmoid activation function. In addition, we have utilized the up-sampled feature maps from 3 different encoding layers including the bottleneck layer. Therefore, we named the model NABLA-3 network. The NABLA-3 model utilizes totally 18.98 Million (M) network parameters. The network is initialized with the He initialization method [21].

## 4. Experiments

### 4.1. Experimental setup

The COVID-19 detection system was developed using multiple models for classification and segmentation tasks which is implemented with TensorFlow deep learning framework on four NVIDIA GTX2080 Ti single GPUs. For testing, the end-to-end system was run on the single GPU system.

### 4.2. Database

We have collected several datasets for implementing this multi-modality learning method. The pneumonia detection samples are collected from publicly available datasets [22]. The total number of samples are 5,216, where only 1,341 samples are for normal and 3,875 samples for pneumonia. The average size of the images is around 1168×984 pixels. Due to the limitation of our computing system, we have resized the images to 128×128 pixels. To resolve the class imbalance problem, we have applied class specific data augmentation. The example images are shown in Figure 5. The COVID-19 dataset has been collected from different sources around the world and a publicly available dataset [23]. Due to the scarcity of training samples, we have applied data augmentation techniques for increasing the number of samples.

For segmentation of the chest regions, we collected 704 chest X-ray images and corresponding masks, which are shown in Figure 5 [24]. The original average size of the sample is 2437×2806×3. We have resized the image to 192×192×3 pixel images. As a result, input samples significantly lose essential information. From the total number of samples, 80% of samples are used for training and the remaining 20% are used for validation and testing of the COVID-Seg_X-ray model. This is the first step of infected regions extraction from input chest X-ray images. The randomly selected X-ray images and corresponding masks are shown in Figure 6.

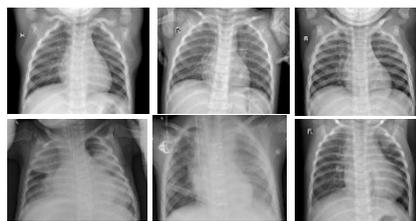

Figure 5: Randomly selected samples without pneumonia in first row and second row shows the X-ray image with pneumonia.

In addition, for COVID-19 detection from CT images, there is no labeled dataset available for this specific task. Hence, we have collected samples from different sources for normal CT scans. The COVID-19 CT images are considered from different CT scans from confirmed patients. A total of 420 samples are collected where 247 samples are for normal and 178 samples are for COVID-19. We have observed the variation in sizes of the samples, ranging from 450×338 pixels to 630×630 pixels.

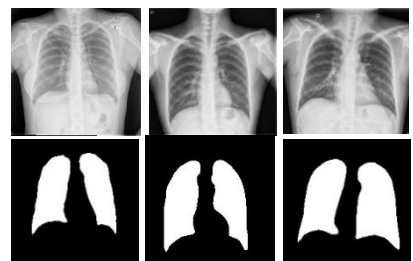

Figure 6: Example images for Chest X-ray segmentation.

Thus, all samples are resized to 192×192 pixels. From the total samples, we randomly selected 375 samples for training and validation, and the remaining 45 samples were used for testing. To increase the number of samples, a data augmentation method was applied during the training. A set of randomly selected samples are shown in Figure 7.

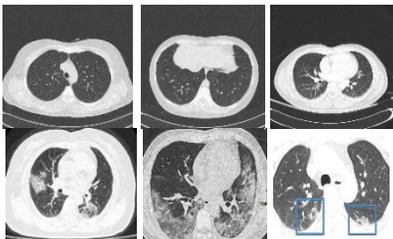

Figure 7: Randomly selected normal and COVID-19 CT images.

For lung segmentation from CT images, we have used a set of publicly available 2D CT scans from Kaggle competition [25]. This dataset contains 267 samples with corresponding masks in total with corresponding labeled images. Some example images and masks are shown in Figure 8.

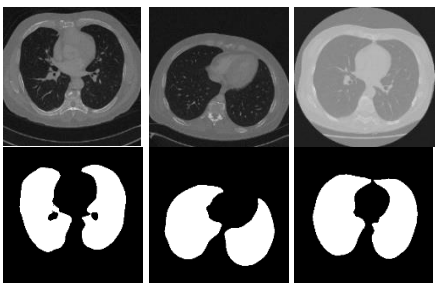

Figure 8: Example images for lung segmentation. First row shows the input images and the second row shows corresponding masks.

The actual size of the images is 512×512. We resized them to 256×256-pixel single channel images. The total number of samples was 267, where 80 percent of the images were used for training and the remaining 20 percent used for validation and testing.

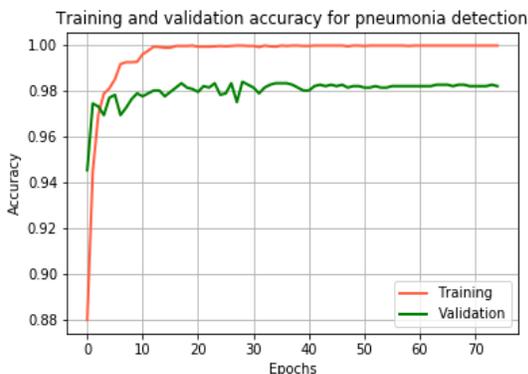

Figure 9: Training and validation accuracy for pneumonia detection task.

### 4.3. Training methods

The IRRCNN model is trained and tested on pneumonia dataset. For COVID-19 detection method for X-ray (COVID-Det_X-ray), the model was trained with the following hyperparameters: Adam optimization method with learning rate $1 \times 10^{-3}$, and Batch size 32. The system was trained for 75 epochs in total where the learning rate was decreased with respect to the factor of 10 after each 25 epochs. The training and validation accuracy for pneumonia detection method is shown in Figure 9. From the figure, it can be seen that the proposed pneumonia detection method shows around 98.2% validation accuracy. COVID-Seg_X-ray was trained with Adam optimizer with learning rate $3\times 10^{-4}$, Dice Coefficient (DC) loss.

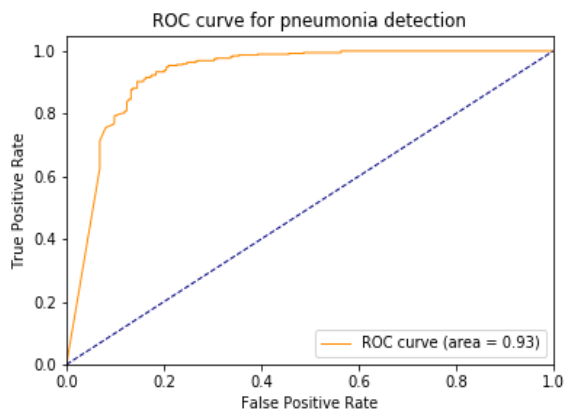

Figure 10: Receiver operating characteristic (ROC)-AUC curve for COVID-19 detection.

The classification model (COVID-Seg_CT) was trained with same parameters as used in COVID-Seg_X-ray, however, due to the lower number of training samples, the model was trained for 150 epochs with batch size of 16. COVID-Seg_CT was trained with Adam optimizer with learning rate $3\times 10^{-4}$, DC loss, As the number of images was less, we have used the batch size of 8. In this implementation, we have used max-min normalized single channel images for training and testing.

### 5. Results

**COVID-Det_X-ray outputs**: After successfully training the model, we have tested the pneumonia detection system with completely new 624 images which includes 234 normal and 390 pneumonia samples. The quantitative results show around 87.26% testing accuracy for pneumonia detection. Then, the same IRRCNN model was used for training and testing for the COVID-19 detection where the pretrained weights from pneumonia were used as initial weights for training the COVID-19 model. We have achieved around 84.67% testing accuracy on the completely new 67 COVID-19 testing samples. The receiver operating

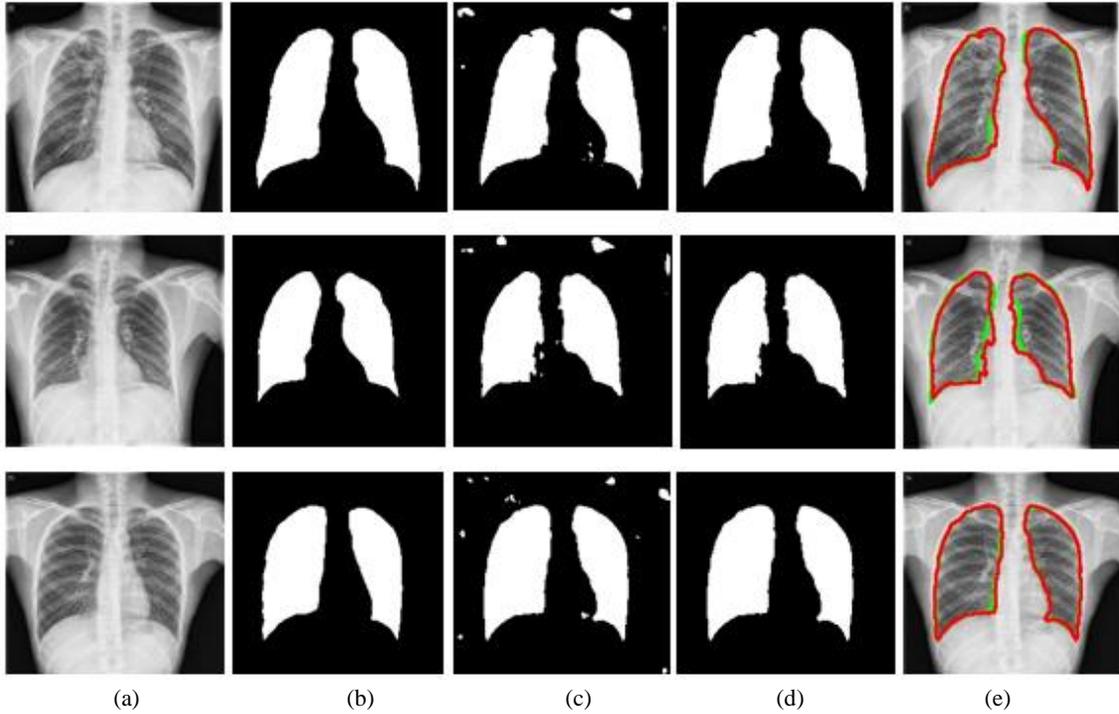
(a) (b) (c) (d) (e)
Figure 11: Chest segmentation (Chest-Seg_X-ray) results: (a) input chest images (b) ground truth (c) model outputs (d) refined outputs and (e) final outputs with contours where the green and red contours represent the ground truth and detection respectively.

characteristics (ROC)-AUC for pneumonia is shown in Figure 10. This shows an AUC of 0.93.

**COVID-Seg_X-ray outputs:** For further evaluation and to define the specific infected regions with COVID-19, we have implemented an end to end segmentation system for extracting the chest regions from the entire X-ray images. The NABLA-N network was used for chest segmentation from X-ray images. This model was trained on the chest-x-ray segmentation challenge dataset. After successfully training the model, the performance was evaluated on the testing samples provided by the organizer. We have tested on completely new 57 chest X-ray images. The model shows 0.9452 and 0.9466 in terms of global accuracy and F1-score respectively. In addition, it shows around 0.8650 and 0.8846 for Intersection over Union (IoU) and Dice similarity Coefficient score (DC).

The experimental results for the segmentation model for COVID-19 X-ray samples are shown in Figure 12. The first column shows the inputs images, the second column shows the outputs from COVID-Seg_X-ray, the third column shows the polished segmentation masks, the fourth column represents outputs for only chest regions, and the fifth column shows the final outputs with COVID-19 infected regions. The qualitative results clearly demonstrate that the proposed model is able to segment and detect contaminated regions of COVID-19 accurately from the chest X-ray images.

In addition, for quantitative justification, we have calculated the total number of pixels of the lung regions and the total number of pixels for infected regions with COVID-19. The percentage is calculated with respect to the total areas of the lung, which can be used for measuring the contingency and severity of corona-virus patients. For the first row and third column in Figure 12, the number of total pixels for lung is 6,696 and total number of infected pixels with COVID-19 is 2,245. Thus, the percentage of infection is 33.52% with respect to the total number of pixels of lung regions. In the last row of Figure 12, the total number of pixels for the lung regions is 9,601 and the total number of infected pixels with COVID-19 is 3,609. Thus the percentage of infection is 37.58%.

This quantitative analysis can be applied to define the severity of the COVID-19 disease. For the second, third and fourth columns, the percentage of infection is around 51.23%, 27.66%, and 47.89% respectively.

**COVID-Seg_X-ray outputs for COVID-19 from Abdominal CT:** We also observed that for some cases, patients have been confirmed with COVID-19 after analysis of the abdominal CT images. Thus, we have evaluated the same model for X-ray as has been tested on abdominal CT images. The outputs for abdominal CT images are shown in Figure 13. The first column shows the input images, the second column represents the refined segmented mask, where only large regions have been selected among a set of segmented regions, and the fourth column shows the extracted lung region with respect to the refined mask shown in Figure 13(b). The classical image processing method and the adaptive thresholding method were applied to extract the pixels to represent COVID-19. The sixth column in Figure 13 demonstrates the heatmaps for the infected regions due to COVID-19. For the first row, the total number of pixels for the lung regions is 5,184 and the total number of infected pixels with COVID-19 is 1,599. Hence, the percentage of infection is 30.84%. For the second and third rows, the percentage of infection is 80.39% and 23.18% respectively. The quantitative and qualitative results clearly show that the proposed classification and segmentation for X-ray images demonstrate promising performance in detection and infected region extraction.

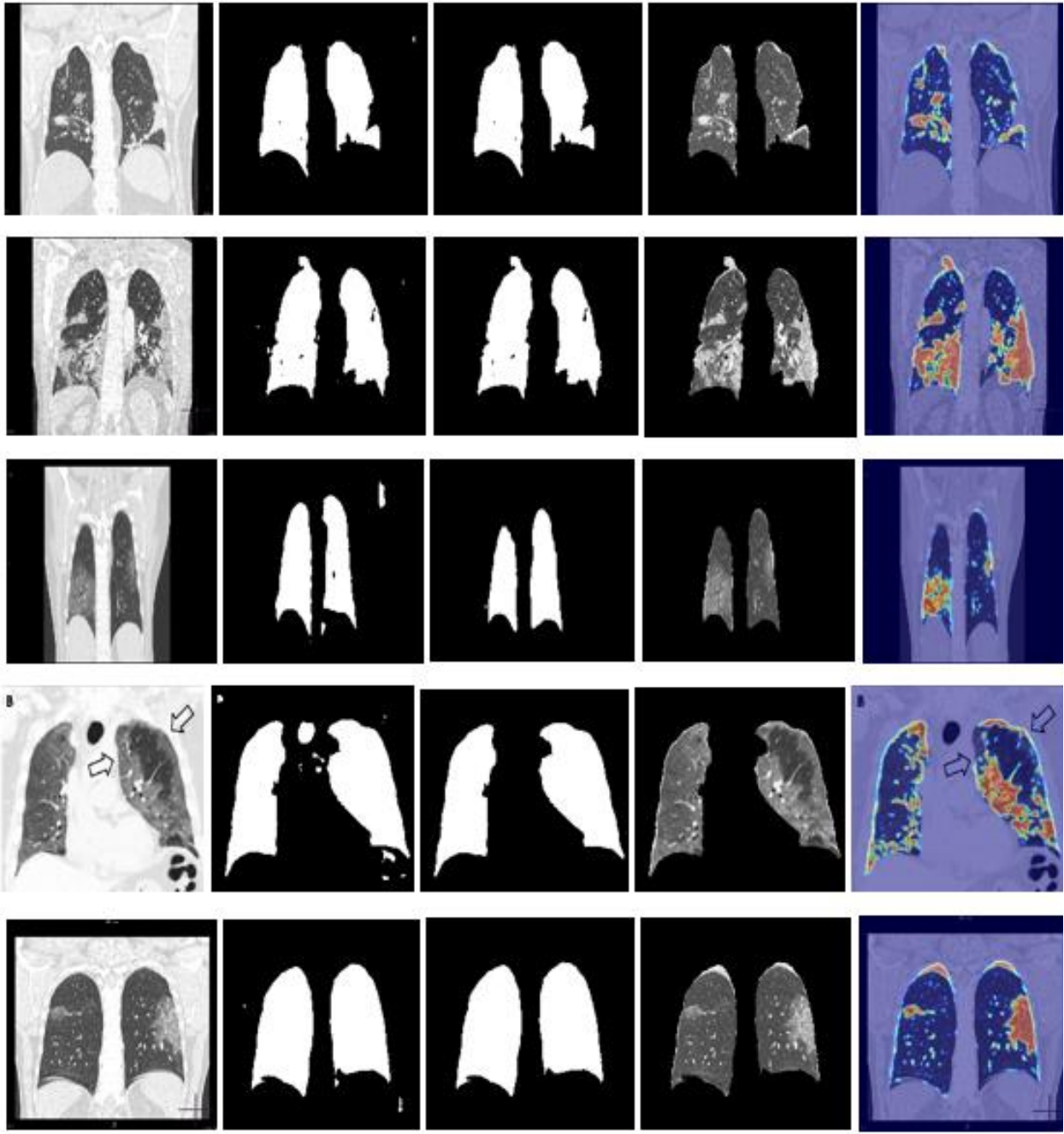

Figure 12: Experimental results of COVID-Seg_X-ray model: first column represents the input images, second column shows the predicted segmentation masks, third column shows the refined outputs with chest regions, fourth column represent only chest regions, and fifth column represent the heatmap in the infected regions.

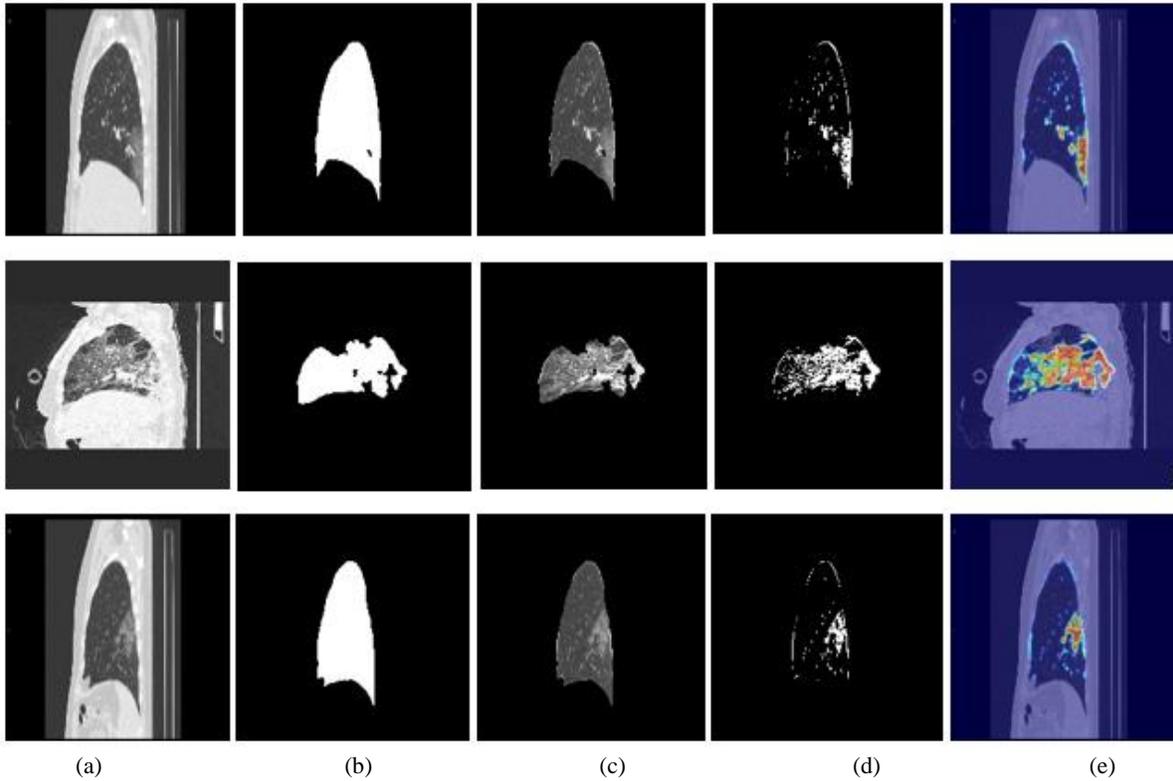

(a) (b) (c) (d) (e)

Figure 13: Experimental results of COVID-Seg_X-ray model for abdominal CT: (a)inputs, (b) segmented masks, (c) lung regions, (d) extracted pixels for COVID-19, and (d)final heat with heatmap.

**COVID-Det_CT outputs:** From the literature survey, we have observed that CT scans are used to identify the COVID-19 detection directly. In some critical cases, Doctors fail to take a decision directly from the X-ray analysis. Instead, the CT scans are used to take final decisions on the patients. Different studies have claimed that CT is more efficient to confirm the COVID-19 patients. Thus, we have included a detection model for CT images which is named COVID-Det_CT. The system is trained and tested on our own dataset. The COVID-Det_CT shows around 98.78% testing accuracy for 45 samples.

**COVID-Seg_CT outputs:** For extracting infected regions with COVID-19, the NABLA-N network-based segmentation model was trained and tested on a publicly available 2D lung segmentation dataset. This model is named COVID-Seg_CT. After training the COVID-Seg_CT model, the testing was done on completely new samples. The quantitative results demonstrate 0.9885 and 0.9956 for F1-score and global accuracy respectively.

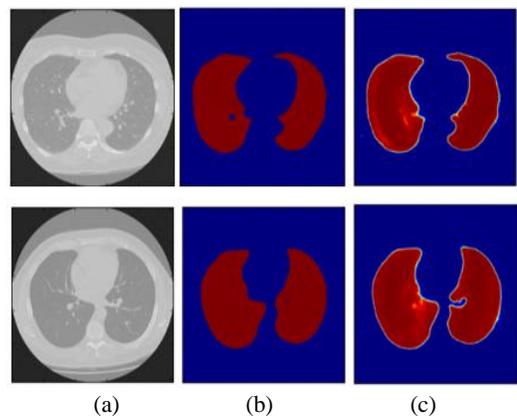

(a) (b) (c)

Figure 14: COVID-Seg_CT outputs for testing samples: (a) inputs (b) Ground Truth (GT) and (c) model outputs.

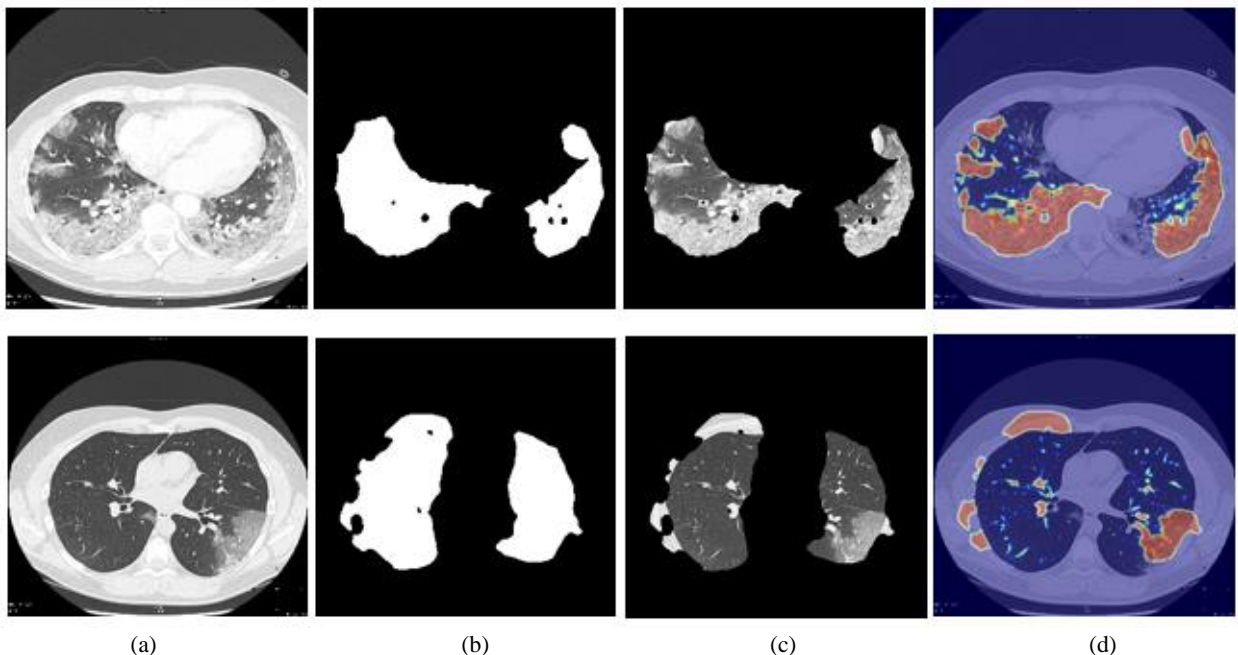

Figure 15: The COVID-19 infected region detection results from lung CT images: (a) inputs images, (b) segmented and refined masks with TL learning approach, and (d) infected region with heatmaps.

The qualitative results are shown in Figure 14. The first column shows the input images, the second column shows the ground truth, and the third column shows the outputs from the COVID-Seg_CT model. The results clearly demonstrate that the proposed model produces very accurate segmentation results compared to ground truth

Since, we did not have any labeled samples for COVID-19 for segmentation tasks, the COVID-Seg_CT model was directly applied to segment the lung regions from the samples with COVID-19. After segmenting the lung region from CT images, mathematical morphological operations were performed to refine the segmentation masks. Lung regions were then extracted with respect to the mask. The segmentation results of COVID-Seg_CT model for COVID-19 samples are shown in Figure 15. The first column shows the input images, the second column shows the lung segmentation results, the third column represents lung regions, and the fourth column represents the results with heatmap. In most of the cases, the proposed model provides good accuracy of detection. However, in some cases, we have observed false detection, as shown in the second row in Figure 15(d).

## 6. Discussions

The proposed pneumonia detection method shows around 87.26% testing accuracy whereas the recently published paper in [15] shows 84.67% testing accuracy. Thus, our IRRCNN based detection model shows around 3.76% better testing accuracy for pneumonia detection tasks. In addition, most of the COVID-19 infected region detection methods proposed are based on patch-based detection methods for infected region extraction, where there is a big possibility to have false positive and false negative detections as the decision is taken based on the class [17]. On the other hand, we proposed an infected region segmentation strategy with pixel level analysis. The qualitative results in our experiments clearly demonstrate efficient detection of infected regions within the lung part and hence the proposed COVID-19 detection significantly reduces the possibility of false positive and false negative detections.

## 7. Limitations

Both detection and segmentation methods for X-ray images provide very promising accuracy. However, there are some limitations of this study which need to be addressed in the near future. First, the COVID-Det_X-ray model needs to be trained and tested with more COVID-19 samples. Second, as the COVID-Det_CT model is trained and tested on only 300 samples in the initial implementation, the model provides very good detection accuracy. From our point of view, the COVID-Det_CT model needs to be trained and tested with more samples to generalize and make the model more robust and accurate. Third, due to the scarcity of the labeled samples for lung segmentation in CT for COVID-19, the COVID-Seg_CT provides outputs with some false positive detections as shown in the second row in Figure 15 (d). This needs to be improved.

## 8. Conclusion

In this study, we proposed an end-to-end system for COVID-19 detection and infected region localization from two different modalities of medical imaging. For classification, and segmentation tasks, our improved Inception Recurrent Residual Neural Network (IRRCNN) and NABLA-3 network models were applied. The models were tested on X-ray, abdominal CT, and full body CT images on publicly available datasets. The observed results show very promising detection results with 84.67% and 98.78% testing accuracy for COVID-19 from X-ray and CT images respectively. In addition, the qualitative results clearly demonstrate high accuracy in the segmentation and detection of infected regions by COVID-19 in both X-ray and CT images. In the near future, we would like to collect more samples of COVID-19 affected subjects to develop a robust and more accurate system.

Classification with NABLA-N and Inception Recurrent Residual Convolutional Networks. Published in SPIE Medical Imaging Conference, 15-20 February 2020, Houston, Texas, USA.

[20] Alom, Md Zahangir, Chris Yakopcic, Mahmudul Hasan, Tarek M. Taha, and Vijayan K. Asari. "Recurrent residual U-Net for medical image segmentation." *Journal of Medical Imaging* 6, no. 1 (2019): 014006.

[21] Alom, Md Zahangir, Tarek M. Taha, Chris Yakopcic, Stefan Westberg, Paheding Sidike, Mst Shamima Nasrin, Mahmudul Hasan, Brian C. Van Essen, Abdul AS Awwal, and Vijayan K. Asari. "A state-of-the-art survey on deep learning theory and architectures." *Electronics* 8, no. 3 (2019): 292.

[22] Mooney. Kaggle chest x-ray images (pneumonia) dataset. https://github.com/ieee8023/covid-chestX-ray-dataset, 2020. 2, 3

[23] Cohen. Covid chest xray dataset. https://github.com/ieee8023/covid-chestxray-dataset, 2020. 2, 3

[24] https://www.kaggle.com/arturscussel/lung-segmentation-and-candidate-points-generation